\documentclass[prl,twocolumn,amsmath,showpacs,floatfix]{revtex4}
\usepackage{graphicx}
\begin{document}

\title{Subcritical finite-amplitude solutions in plane Couette flow of visco-elastic fluids}

\author{Alexander N. Morozov}
\author{Wim van Saarloos}

\affiliation{Instituut-Lorentz,
Universiteit Leiden, Postbus 9506, 2300 RA Leiden, The Netherlands}

\date{\today}

\begin{abstract}
Plane Couette flow of visco-elastic fluids is shown to exhibit a purely elastic subcritical instability in spite of
being linearly stable. The mechanism of this instability is proposed and the nonlinear stability analysis of plane Couette
flow of the Upper-Convected Maxwell fluid is presented. It is found that above the critical Weissenberg number, a small 
finite-size perturbation is sufficient to create a secondary flow, and the threshold value for the amplitude of the 
perturbation decreases as the Weissenberg number increases. The results suggest a scenario for weakly turbulent 
visco-elastic flow which is similar to the one for  Newtonian fluids as a function of Reynolds number.

\end{abstract}


\maketitle

Recently, it has been discovered that flows of visco-elastic fluids like polymer solutions and melts can lose their stability
and become turbulent at very low Reynolds numbers \cite{Larson_Shaqfeh_Muller,McKinley_1991,
Groisman_Steinberg_Nature1,Groisman_Steinberg_2004}. 
In contrast to the Newtonian turbulence,
where inertia plays a destabilizing role, this {\it elastic turbulence} or {\it turbulence without
inertia} \cite{Larson_Nature} arises when the flow-induced stretching of the polymers makes the elastic stresses in the fluid become large and anisotropic.
It is a challenge to explain how the transition to this new type of turbulence can occur.
In this Letter we show that visco-elastic plane Couette flow (PCF) exhibits a subcritical instability to finite amplitude
waves and argue for a scenario for weak elastic turbulence analogous to that for weak turbulence in PCF of Newtonian fluids.

The mechanism of linear elastic instability was identified for flows with curved stream-lines \cite{Larson_Shaqfeh_Muller,Shaqfeh_1992}. One 
of the classical examples of such a flow is realized in Taylor-Couette cell where fluid 
fills the gap between two coaxial cylinders made to rotate with respect to each other. In the laminar
state, the fluid moves around the cylinder axis and the elastic or {\it hoop} stresses act on 
polymer molecules stretching them along the circular stream-lines and exerting extra pressure 
towards the inner cylinder. When these stresses overcome viscous friction, the laminar state
becomes linearly unstable - any infinitesimal perturbation will push a polymer from the
circular stream-lines and create a secondary flow, the so-called Taylor vortices. 
Pakdel and McKinley generalized this mechanism to arbitrary flows \cite{McKinley_PRL} and proposed that there exists a universal
relation between the properties of the fluid and the flow geometry which determines the conditions of
the linear instability. One of the dimensionless parameters in their argument is the so-called {\it Weissenberg number}
$Wi=\lambda\, \dot{\gamma}$, where $\lambda$ is the elastic relaxation time of the fluid, and the shear rate $\dot{\gamma}=\partial v / \partial x$
gives the relative velocity of two fluid layers moving with respect to each other.
Pakdel and McKinley argued \cite{McKinley_PRL} that the critical Weissenberg number is related to the characteristic
curvature of the flow stream-lines and that the linear instability disappears when the curvature goes to zero.
The known results on the visco-elastic instabilities in Taylor-Couette \cite{Larson_Shaqfeh_Muller}, cone-and-plate and
parallel plate \cite{McKinley_1991}, Dean and Taylor-Dean \cite{Shaqfeh_1992} flows are in agreement with this 
{\it curved stream-lines - linear instability} paradigm.

The linear stability of parallel visco-elastic shear flows has been investigated in detail. For essentially all studied visco-elastic models, laminar PCF is linearly stable \cite{Gorodtsov_Leonov,Renardy} (note the exception \cite{Grillet}). 
In the case of the pipe flow, the linear
stability was demonstrated numerically by Ho and Denn \cite{Ho_Denn} for any value of the Weissenberg and Reynolds numbers. 
Therefore, it has become common knowledge that the parallel shear flows of fluids obeying simple visco-elastic models (UCM, Oldroyd-B, etc.)
are linearly stable, in agreement with the {\it curved stream-lines - linear instability} paradigm. Clearly, if an instability does occur in practice, it  has to be nonlinear.

There
is no general agreement, however, on whether parallel shear flows like PCF or pipe do in fact become unstable. At the moment, there has been no
experiment that would clearly establish the presence or absence of a bulk hydrodynamic instability in parallel visco-elastic shear flows. One of a few
indirect indications that a bulk instability might occur in the pipe flow comes from the
famous melt-fracture problem \cite{Denn_reviews}, which arises in extrusion of a dense polymer solution or melt through
a thin capillary. There, when the extrusion rate exceeds some critical  value,
the surface of the extrudate becomes distorted and the extrudate might even break, giving the name to the
phenomenon. It is possible that this is a manifestation of an instability taking place inside the capillary,
though other mechanisms (such as stick-slip, influence of the inlet, etc.) have been proposed \cite{Denn_reviews}.
Recently, we presented arguments for the bulk instability being related to the melt-fracture phenomenon 
\cite{Meltfracture_us}, but the
issue stays highly controversial. There is also some evidence for nonlinear parallel shear flow instabilities
from numerical simulations of visco-elastic hydrodynamic equations \cite{Atalik_Keunings}. Partly because the numerical schemes used to solve these
equations are known to break down when elastic stresses become large ($Wi \gtrsim 1$) --- the so-called {\it large
Weissenberg number problem} \cite{Owens_Phillips} --- it is open to debate whether an observed phenomenon is due to a numerical or a true physical instability.

In this Letter we show explicitly that  a nonlinear instability mechanism does exist in agreement with the following argument. The laminar
velocity profiles of the parallel shear flows have straight stream-lines, and, therefore, their linear stability is in agreement with
the {\it curved stream-lines - linear instability} paradigm. The linear theory predicts that a small perturbation superimposed
on top of the laminar flow will decay in time with the decay rate depending on the Weissenberg number. When $Wi$ becomes larger
than one, the decay time becomes comparable with the elastic relaxation time $\lambda$, and the perturbation becomes long-living.
Thus, on short time-scales, the superposition of the laminar flow and the slowly-decaying perturbation can be viewed as a new
basis profile {\it with} curved stream-lines. Applying the same {\it curved stream-lines - linear instability} paradigm to the perturbed streamlines, we conclude
that this new flow can become linearly unstable. The instability occurs via a perturbation on top of a perturbation, and thus is
nonlinear. Since the initial perturbation has to be strong enough to become unstable, there exists a finite-amplitude
threshold for the transition, which becomes smaller as the Weissenberg number increases. This scenario resembles transition to turbulence
in parallel shear flows of Newtonian fluids. There as well, one encounters the absence of the linear instability, and a subcritical transition with the
threshold going down with the Reynolds number \cite{Henningson_book,Bjorn_Hof_PRL}.

Our explicit results are for the nonlinear stability analysis of PCF. 
We consider the so-called UCM fluid \cite{Bird}  confined in the $y$-direction in-between two plates ($y=\pm d$), which move with constant
velocity $v_0$ in the opposite directions along the $x$-axis. The hydrodynamic equations, consisting of the equations for momentum balance, 
the UCM model, and incompressibility, 
read \cite{Bird}
\begin{eqnarray}
\label{ucm1}
&&\qquad Re \left[ \frac{\partial {\mathbf v}}{\partial t} + \left({\mathbf v}\cdot{\mathbf \nabla}\right){\mathbf v}\right] = -{\mathbf \nabla} p 
- {\mathbf \nabla}\cdot{\pmb{\tau}}, \\
&&{\pmb{\tau}} + Wi \left[ \frac{\partial{\pmb\tau}}{\partial t} + {\mathbf v}\cdot{\mathbf \nabla}{\pmb\tau} - \left(\nabla{\mathbf v}\right)^\dagger\cdot{\pmb\tau} 
  - {\pmb\tau}\cdot(\nabla\mathbf{v})\right] \nonumber \\
\label{ucm2}
&&\qquad\qquad\qquad\qquad \qquad= -\left[ (\nabla\mathbf{v})+(\nabla\mathbf{v})^\dagger \right],\\
\label{ucm3}
&&\qquad\qquad\qquad\qquad{\mathbf{\nabla}\cdot{\mathbf{v}}}=0 ,
\end{eqnarray}
where the Weissenberg number $Wi$$=$$\lambda v_0/d$, the Reynolds number $Re=\rho v_0^2 / (\eta d)$, $\rho$ is the density and $\eta$ is the viscosity of the
fluid, $\pmb\tau$ and $\mathbf v$ are the stress tensor and the velocity, respectively; $d$ is used as the unit of length, $d/v_0$ as the unit of time, 
and the stress tensor is scaled with $\eta v_0/d$; $(...)^\dagger$ denotes the transposed matrix. As usual, 
we split these variables in two parts: the laminar values $\{{\pmb \tau}_{lam},y\,{\mathbf e_x}\}$ and the deviation describing the disturbance 
$\{{\pmb \tau}',{\mathbf v}'\}$.
It is useful to organize all hydrodynamic fields of the disturbance in one vector $V=\{\tau'_{ij},v'_i,p\}^\dagger$. 
Then, the equations (\ref{ucm1}-\ref{ucm3}) can  formally be
re-written as
\begin{equation}
\label{operator_equation}
\hat{\mathcal{L}}\,V + \frac{\partial}{\partial t} \{ Re\,\,\,{\mathbf v}', Wi\,\,\,{\pmb\tau}',0 \}^\dagger = N\left(V,V\right) ,
\end{equation}
where the l.h.s. represents the linear terms in eqs.(\ref{ucm1}-\ref{ucm3}), and the r.h.s. the quadratic nonlinearity.

The first step of our analysis is to determine the eigenvalues $\lambda$ and the eigenfunctions $V_0$ of the linear operator $\hat{\mathcal{L}}$. 
Gorodtsov and Leonov \cite{Gorodtsov_Leonov},
and Reynardy {\it et al.} \cite{Renardy} have shown that the eigenfunctions in the form
\begin{equation}
V_0(x,y,z)=\tilde{V}_0(y) e^{i(k\,x+q\,z)} + {\mathit c.c.}
\end{equation}
have two types of physical eigenvalues: a pair of complex-conjugated "elastic" eigenvalues (Gorodtsov-Leonov modes), and an infinite discrete set of "inertial" ones.
(There also exists a continuous spectrum of eigenvalues which were shown to be unphysical \cite{Graham_JFM_1998} and which, therefore, will be discarded.)
Let us for the moment focus on the elastic mode,  and suppose we choose the initial disturbance to be in the form
of the elastic eigenmode
\begin{equation}
\label{solution_form}
V(x,y,z,t)=\Phi(t)\,\tilde{V}_0^{(GL)}(y) e^{i(k\,x+q\,z)} + {\mathit c.c.}
\end{equation}
where $\Phi(t)$ is a complex amplitude
(Our normalization is such that when the amplitude $\Phi(t)=1$, the strength of the shear rate created by the perturbation equals that of the laminar flow). In order to investigate
how  the amplitude $\Phi(t)$ will change in time depending on $Wi$, $k$ and $q$, we derive the equation governing the time evolution of
$\Phi(t)$, or {\it the amplitude equation}. The standard technique used to derive the amplitude equations relies 
on the presence of linear instability (occurring at,
say, $Wi_{\rm lin}$), and uses the distance to the instability $(Wi-Wi_{\rm lin})/Wi_{\rm lin}$ as a small parameter. Then, the nonlinear evolution of $\Phi(t)$ near $Wi_{\rm lin}$
can be deduced by the method of multiple scales \cite{Cross_Hohenberg}. The parallel shear flows, however, are linearly stable ($Wi_{\rm lin}\rightarrow\infty$, effectively), 
and we
have to use other methods. Instead, assuming $\Phi(t)$ to be small, we substitute Eq.~(\ref{solution_form}) into Eq.~(\ref{operator_equation}), collect the terms
proportional to $\exp{\left[i (k x + q z)\right]}$, and using the corresponding eigenmode of the adjoint operator 
$\hat{\mathcal{L}}^\dagger$, we project these terms on the original form Eq.~(\ref{solution_form}). We then find the time derivative of $\Phi(t)$ as a series in $\Phi(t)$
\begin{eqnarray}
&&\frac{d\Phi}{dt}=\lambda^{(GL)}\Phi + C_3 \Phi |\Phi|^2 + C_5 \Phi |\Phi|^4 + C_7 \Phi |\Phi|^6 \nonumber \\
\label{amplitude_equation}
&& \qquad \qquad +\,C_9 \Phi |\Phi|^8 + C_{11} \Phi |\Phi|^{10} + \cdots
\end{eqnarray}
where $\lambda^{(GL)}$ is the eigenvalue of the elastic (Gorodtsov-Leonov) mode, and the nonlinear coefficients $C$'s are explicit functions of $Wi$, $k$ and $q$. 
Eq.~(\ref{amplitude_equation}) has solution $\Phi(t)=|\Phi| e^{i\,\Omega\,t}$ which results in traveling waves in Eq.~(\ref{solution_form}). 
For small amplitudes, it reproduces the linear decay $\Phi\sim\exp\left[\lambda^{(GL)}t\right]$, $Re\left( \lambda^{(GL)}\right)<0$.
The instability
threshold is determined by finding the steady-state solutions of eq.(\ref{amplitude_equation}). The main problem in dealing with the series like eq.(\ref{amplitude_equation})
is that it is not known {\it a priori} whether it converges. 
In order to check the convergence of the series upon inclusion of higher-order terms, we solve the equation ${\rm Re}\left(\frac{d\Phi}{dt}\right)=0$ in successive orders.

\begin{figure}
\begin{center}
  \includegraphics[width=0.81 \linewidth]{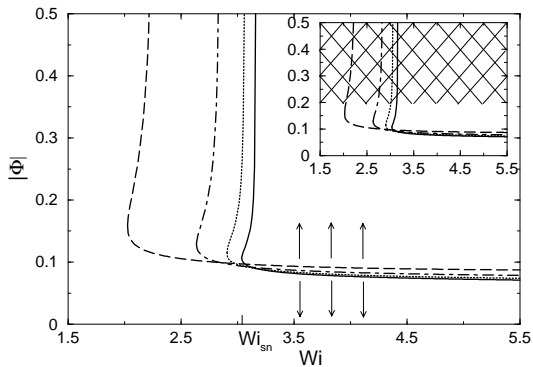}
\end{center}
\vspace*{-0.4cm}

\caption{Steady-state amplitude $\Phi$ for $k=1$ and $q=1$: dashed, dot-dashed, dotted and solid lines show the solution to $Re\left(\frac{d\Phi}{dt}\right)=0$ 
up to the $5^{th}$, $7^{th}$, $9^{th}$ and $11^{th}$ order.
The inset illustrates qualitatively that the series converges only for small enough $|\Phi|$.
The ratio $Re/Wi=10^{-3}$ was kept constant.}
\label{PC}
\end{figure}

In Fig.\ref{PC} we show the solution to these equations for $k=1$ and $q=1$. The most important feature of these curves is that they show existence
of a subcritical instability for Weissenberg numbers larger than the saddle-node value $Wi_{\rm sn}$ indicated in the figure and obtained by extrapolating the curves. As the arrows indicate, for $Wi>Wi_{\rm sn}$ the lower branch of the curves denotes the critical amplitude --- amplitudes larger than this value will grow in time. 
Note that  the instability threshold is small 
(consistent with our assumption $|\Phi|<1$), and goes down as $Wi$ increases. The inclusion of higher-order terms causes the whole curve to shift 
to the right, though the shift becomes roughly two times smaller with every coefficient included, indicating convergence.

While the lower branch of each curves gives the minimal amplitude of the disturbance sufficient to destabilize the laminar flow, the upper branch determines
the saturated value of $\Phi$ after the transition. Surprisingly, it diverges in the vicinity of the saddle-node where the highest coefficient in 
the expansion changes sign. There could be several reasons for
that. First, it may indicate that the nonlinear state in the form of Eq.~(\ref{solution_form}) is unstable and will undergo a transition to another coherent 
state or to turbulence. Second, 
although we consider this unlikely, it may be
that the UCM model cannot capture this state. Finally, it may be a convergence problem. As indicated in the inset of Fig.\ref{PC}, the upper branch may 
lie beyond the radius of convergence of 
(\ref{amplitude_equation}). When we apply the method to the subcritical Swift-Hohenberg equation \cite{Paul_in_preparation}, we also find such behavior.

\begin{figure}
\begin{center}
  \includegraphics[width=0.81 \linewidth]{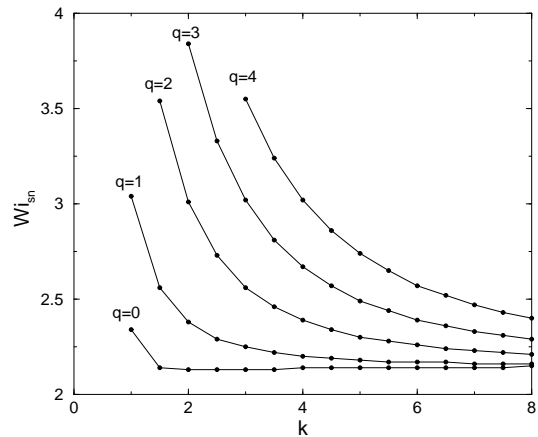}
\end{center}
\vspace*{-0.4cm}

\caption{The saddle-node Weissenber number as a function of the wave-vectors. The dots are the calculated values, the lines serve as a guide to the eye.}
\label{saddle-node}
\end{figure}

In Fig.\ref{saddle-node} we plot the lowest Weissenberg number for which the nonlinear instability is possible, or the position of the
saddle-node $Wi_{\rm sn}$, as a function of the wave-vectors $k$ and $q$. It clearly shows that the saddle-node position is only a weak function
of the wave-vectors, and a large number of modes with different $k$'s and $q$'s is nonlinearly unstable for given $Wi>2.1$. Even if each
individual mode saturates at a given value of $\Phi$, the superposition of a large number of such modes is likely to become chaotic, and we
expect the elastic turbulence to set 
close to or even at the instability.

The existence of the subcritical instability in PCF can also be deduced from the well-studied Taylor-Couette flow 
\cite{Larson_Shaqfeh_Muller,Shaqfeh_1992,Avgousti_1993}. 
There, the subcritical
character of the transition was both observed experimentally \cite{Groisman_Steinberg_PRL_1997} and calculated analytically 
\cite{Sureshkumar_1994}. We have performed the nonlinear stability
analysis for the purely elastic eigenvalue of the non-axisymmetric mode used by Sureshkumar {\it et al.} \cite{Sureshkumar_1994}, which we have traced 
to the case of the counter-rotating cylinders. We have found that this mode becomes subcritically unstable far below the linear instability threshold.
Moreover, as the curvature of the cylinders goes to zero, the linear instability disappears at infinity, while the threshold of the subcritical
transition goes smoothly to the PCF one calculated above. This is somewhat similar to the Nagata's construction for the Newtonian
fluids \cite{Nagata_1990}. The full report on these results will be published elsewhere.

Now we turn to the discussion of the "inertial" eigenmodes. Replacing $\lambda^{(GL)}$ and $\tilde{V}_0^{(GL)}$ in Eq.~(\ref{solution_form}) with one of
the inertial eigenvalues and the corresponding eigenmode, and repeating the same calculation for the coefficients in the amplitude equation 
(\ref{amplitude_equation}), we find that this type of disturbances is nonlinearly stable for $Wi\sim Wi_{\rm sn}$. We have checked  this result for
the three eigenvalues with the smallest imaginary parts. The structure of the eigenspectrum also helps to rule out the possibility that the
disturbances in the form of the inertial eigenmodes can become unstable via a nonlinear interaction with the elastic mode. To the lowest order, the
interaction between the elastic amplitude $\Phi$ and an inertial amplitude $\Psi$ is described by adding terms like $C_3^{(\Phi\Psi)}\Phi |\Psi|^2$ to
the amplitude equation (\ref{amplitude_equation}), and similar terms to the equation for $\Psi$. Since
the imaginary part of the elastic eigenvalue is $O(1)$, while the imaginary part of the inertial eigenvalues is $O(1/\sqrt{Re}) \gg 1$ for small $Re$,
the coefficient $C_3^{(\Phi\Psi)}$, describing 
the overlap between two modes, must be very small, so the interaction terms can be 
neglected.

The above results give very strong indications for the existence of a branch of nonlinear finite-amplitude  solutions which renders visco-elastic PCF flow nonlinearly unstable for $Wi \gtrsim 2.1$. We believe these solutions do play an important role in organizing the dynamics of visco-elastic PCF. First of all, it is intriguing to note that in the direct numerical simulations of PCF of Atalik and Keunings \cite{Atalik_Keunings}, numerical instabilities were found for $Wi\gtrsim 2$ --- could this be due to the occurrence of these modes? Secondly, we have found that the branch of solutions which we have established here have an analog in  the pipe flow  where the transition is found to be at $Wi_{\rm sn}\approx 5$ (results will be published elsewhere, see also \cite{Meltfracture_us}). Thirdly, and most importantly, the similarity of our results with what has been found for PCF of Newtonian fluids lead us to suspect that these solutions could be the building blocks of weak visco-elastic turbulence: The visco-elastic periodic finite-amplitude solutions found here resemble the periodic finite-amplitude solutions in Newtonian fluids. There, they are known to be
{\it exact} but {\it unstable} solutions of the Navier-Stokes equation \cite{Nagata_1990} and take part in the self-sustaining cycle --- a
periodic sequence of instabilities in which streaks, stream-wise vortices and rolls are continuously destroyed and regenerated \cite{Waleffe_1997,Bjorn_Hof_Science}.
It is tantalizing to speculate that a similar cycle can be proposed to sustain the weak elastic turbulence in PCF and Poiseuille flow. 
The first
step in this direction was made in \cite{Stone} where the influence of a minute amount of polymer on the weak Newtonian turbulence was studied.
Together with our findings on the finite-amplitude solutions this gives hope that a visco-elastic version of the self-sustaining cycle can indeed be formulated.

In conclusion, we have presented evidence for nonlinear instability in visco-elastic PCF. Together with the vast experimental
and numerical evidence in various flow geometries 
\cite{Groisman_Steinberg_2004,Groisman_Steinberg_PRL_1997,Kumar_Graham_PRL,McKinley_1991,Meltfracture_us,Sureshkumar_1994}, 
this makes us believe that the nonlinear subcritical instability is an inherent feature of the visco-elastic parallel shear flows and 
that the finite-amplitude solutions may organize dynamics of weak elastic turbulence.

\end{document}